# Far-field Compound Super-resolution Lens for direct laser writing of arbitrary nano patterns and beyond

B. Yan,[1,2] L. Yue,[1] J. N. Monks,[1] X. Yang, [2,b)] D. Xiong,[2] C. Jiang,[3] Z. Wang, [1,a)]

[1] School of Computer Science and Electronic Engineering, Bangor University, Dean Street, Bangor, Gwynedd, LL57 1UT, UK

[2] Center of Optics Health, Suzhou Institute of Biomedical Engineering and Technology, Chinese Academy of Sciences, No. 88 Keling Street, Suzhou Jiangsu, 215163, China

[3] College of Electrical and Information Engineering, Northeast Petroleum University, Daqing, 163318, China

**Corresponding Author**

a) E-mail (Z. Wang): z.wang@bangor.ac.uk

b) E-mail (X. Yang): yangxb@sibet.ac.cn

**ABSTRACT**:

A low-cost compound super-resolution lens, consisting of a Plano-Convex lens and a Microsphere lens (PCM), was proposed and demonstrated for subwavelength direct laser scanning writing application. The PCM lens can achieve a far-field super-resolution of $\sim\lambda/3.5$ in air ($\sim 6$ μm away for lens) which surpasses resolution limit of existing commercial objective lenses and is the first of its kind. Arbitrary nano-patterns can now be directly fabricated on various substrates in a simple and low-cost manner using developed PCM lens. The lens can also be used for other applications including nano-imaging and sensing as well in a confocal configuration. This work may lead to the development of next-generation low-cost direct laser nanofabrication machine and super-resolution imaging nanoscope.

**Introduction**

Laser has been recognized as one of the most extensively used tools for micro/nano-patterning. Complicated structures can be precisely generated through a noncontact and maskless laser direct-writing. However, the key challenge of laser processing to produce extremely small features is the optical diffraction limit.[1] To overcome such difficulty, various laser-based techniques were emerged for sub-diffraction nano-texturing, such as nearfield scanning optical microscope (NSOM) patterning,[2,3] laser combination with scanning probe microscopy (SPM) tip patterning,[2,4] plasmonic lithography,[5] laser thermal lithography (ITL),[6] interference lithography[7,8] as well as multiphoton absorption lithography,[9,10] etc. However, these techniques were limited in laboratory stage due to their low throughput and sophisticated control system.

Recently, using dielectric microsphere as a nearfield lens for super-resolution nano imaging and fabrication has attracted great interests. The optical phenomenon knowns as photonic nanojet can contribute to laser beam focusing to break the diffraction limit.[11-13] Researchers have achieved sub-diffraction features (~$\lambda$/3) through laser-induced particle lens technique.[14,15] In order to increase the throughput, contacting particle lens arrays (CPLA) technique was also introduced. This method employs a monolayer close-packed particle-lens array to split laser beam into multiple enhanced focal spots to generate parallel nano-features over large area.[16-18] Guo et al., on this basis, has improved and innovated a kind of arbitrary-shaped patterning technique by an adjustable angular incident laser beam.[19]

Although CPLA has been proven to be a simple and highly productive strategy for nano-fabrication on various materials, there are some inevitable limitations. The target substrate for preparing preliminary self-assembly monolayer has to be hydrophilic which implies the inadaptability of hydrophobic surface.[20] Furthermore, the ejection of microspheres generally happens after a single laser shot, which may due to thermal expansion of substrate caused by laser absorption.[21] Therefore, the substrate dependency and non-reusable factor make it impossible for industry use. Several techniques were proposed to circumvent these issues, usually by transferring CPLA to transparent host media.[22-24] Although multi-pulse processing is feasible by this method, the series of techniques are still lack of reliability and cannot produce user-defined arbitrary nanostructures. Therefore, there is a strong need to further develop this technique to meet industry processing requirement.

In order to realize accurate and smooth scan patterning process, a gap between focusing lens and sample is a necessary condition. Contrariwise, contact-mode may result in dragging issues due to lens-sample friction, which can lead distortion of final patterns. In addition, unintentional scratches on delicate samples may also be generated. Most researches have studied the microsphere lens patterning technique based on nearfield mode that is contact or within an incident wavelength distance. Recently, manipulation of single microsphere lens by laser trapping and tip-based scanning techniques were demonstrated for complex nano-pattern processing.[25,26] There were both worked in the near-field modes, either limited to liquid environment (laser trapping) or involving sophisticated and costly near-field tip control system. We wish to develop a far-field, low-cost, super-resolution microsphere scanning writing system for direct laser writing of arbitrary nano patterns, which can be further extended to super-resolution imaging and sensing applications.

In this paper, we report a major step forward in laser direct patterning system that can perform large-area subwavelength surface processing with user-defined patterns. A new superlens design, namely a compound lens consisting of a Plano-Convex lens and a Microsphere lens (PCM), was proposed. The PCM superlens can be implemented as an optical probe to generate subwavelength focusing at micrometer distance. With the assist of side-view monitoring system and high-resolution nano-stage, we can achieve precise control and monitor of the working distance between probe and sample during scan patterning process. Meanwhile, the performance of PCM lens at different working distance was theoretically and experimental evaluated. The capability of arbitrary pattern fabrication was realized. It is an objective-free, low-cost, high efficient and flexible system with high completion for large-area sub-wavelength nano-patterning, which will satisfy the growing industrial demands in laser nano-patterning.

**Experimental Setup**

The experimental setup is illustrated in Figure 1a and c. The supporting frame was made of 3D printed structures and metal struts to install lens and stages. A Barium Titanium Glass (BTG) microsphere with diameter of 50 μm was aligned and attached onto PDMS SYLGARD 184, DOW CORNING) or UV glue (NOA63, Noland Adhesive) pre-coated curved surface of a BK7 Plano-Convex lens (LA1700, Thorlabs). A second thin layer of PDMS or UV glue was then applied to form partially encapsulation of BTG microsphere by spin-coating at 2000 rpm for 1 minute (Figure 1d). This results in the formation of a probe-like Plano-Convex microsphere (PCM) lens, where a single microsphere slightly extruding out (Figure 1e). Afterwards, the PCM lens was bonded onto a 3D printed lens holder to build a detachable PCM lens module (Figure 1b). A manual 3-axis stage was used for coarsely adjusting target sample. A long working distance zoom lens was placed horizontally at side for monitoring the gap between microsphere tip and samples. The light sources were generated by femtosecond laser consisting of a Ti:Sapphire oscillator (Spectra Physics Tsunami) and a regenerative amplifier (Spectra Physics Spitfire) which provides 800 nm wavelength, 100 fs pulse duration and 5 KHz repetition rate. The beam diameter projected at the PCM lens incident plane was ~200 μm.

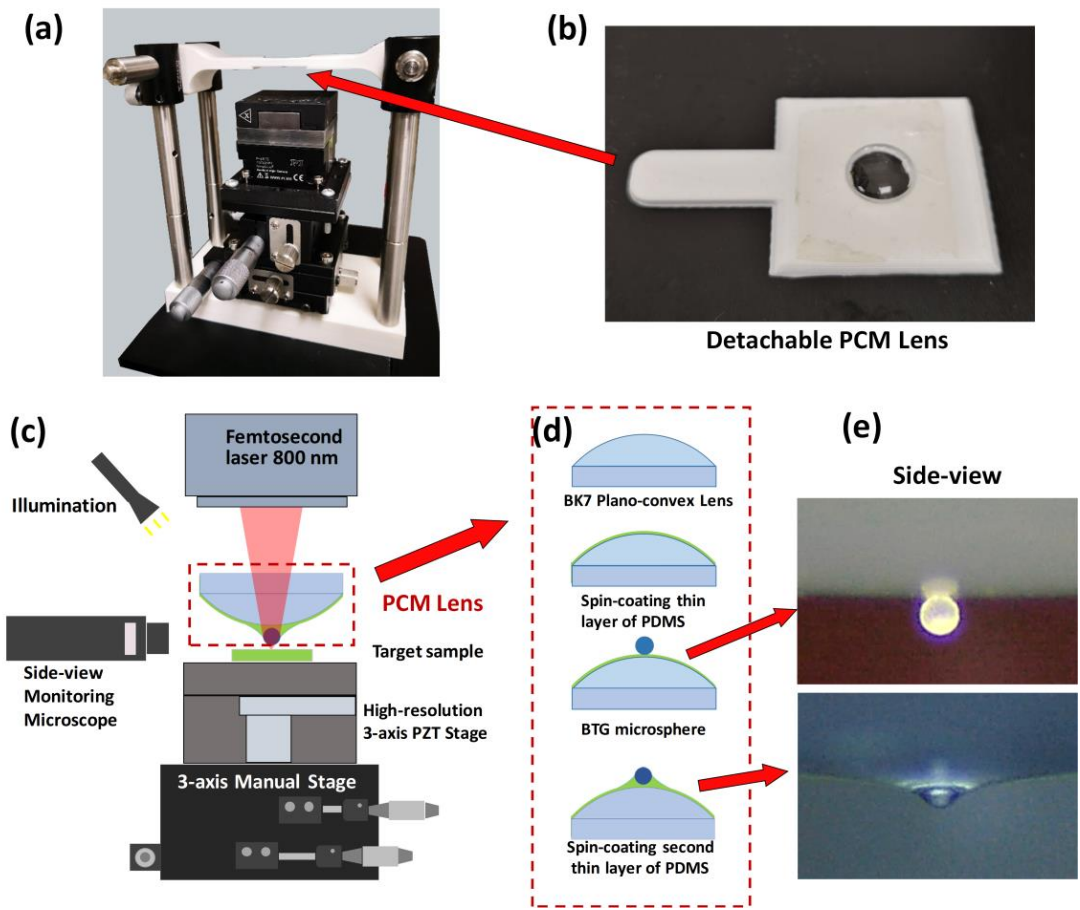

**Figure 1.** PCM lens and system setup for laser nano-patterning. (a) Photograph of actual system setup. (b) Photograph of actual detachable PCM lens module. (c) Schematic of the PCM nano-patterning system. (d) Fabrication of PCM lens. (e) Side view of PCM lens.

The scanning was performed using a high-resolution nano-stage (P-611.3 NanoCube, Physik Instrumente), with 1 nm resolution in the XYZ direction, and a travel range of 100 μm. In our experiments, the PCM lens was kept static and the underlying nano-stage moved and scanned the samples across the PCM lens. The laser processing was automatically undertaken at the synchronization of laser and nano-stage through in-house developed GCS code. Samples including silicon wafer, nickel and Au thin film coated glass were tested.

**Results and Discussion**

To understand focusing characteristics of the PCM lens, computational modelling was performed using in-house developed code based on physical optics. A planar wave (800 nm) propagates through a 50 μm Barium Titanate Glass (BTG) microsphere ($n_p$=1.9) partially encapsulated by PDMS material ($n_m$=1.4) to imitate PCM lens in the experiment. Figure 2a shows the calculated optical field distribution in XZ-plane. It reveals that the electric field exhibits elongated shape and diverges with distance. The field maximum was found at a 9.14 μm distance away from microsphere bottom boundary and intensity drops quickly onwards. Figure 2b illustrates the field profiles along y direction at different working distances. We observed multi-peak focusing phenomena before the light convergence and significant side-peak occurs around 7 μm distance. The focal spot size with respect to the distances was revealed in Figure 2c. The calculated full-with at half-maximum (FWHM) starts from 200 nm at boundary of microsphere, then gradually arises to 670 nm as the distance increases to 11 μm, afterwards it reaches at micro-scale due to light divergence. The super-resolution ($\lambda/2<400$ nm) is found for distances within 6 μm measured from the bottom of PCM lens.

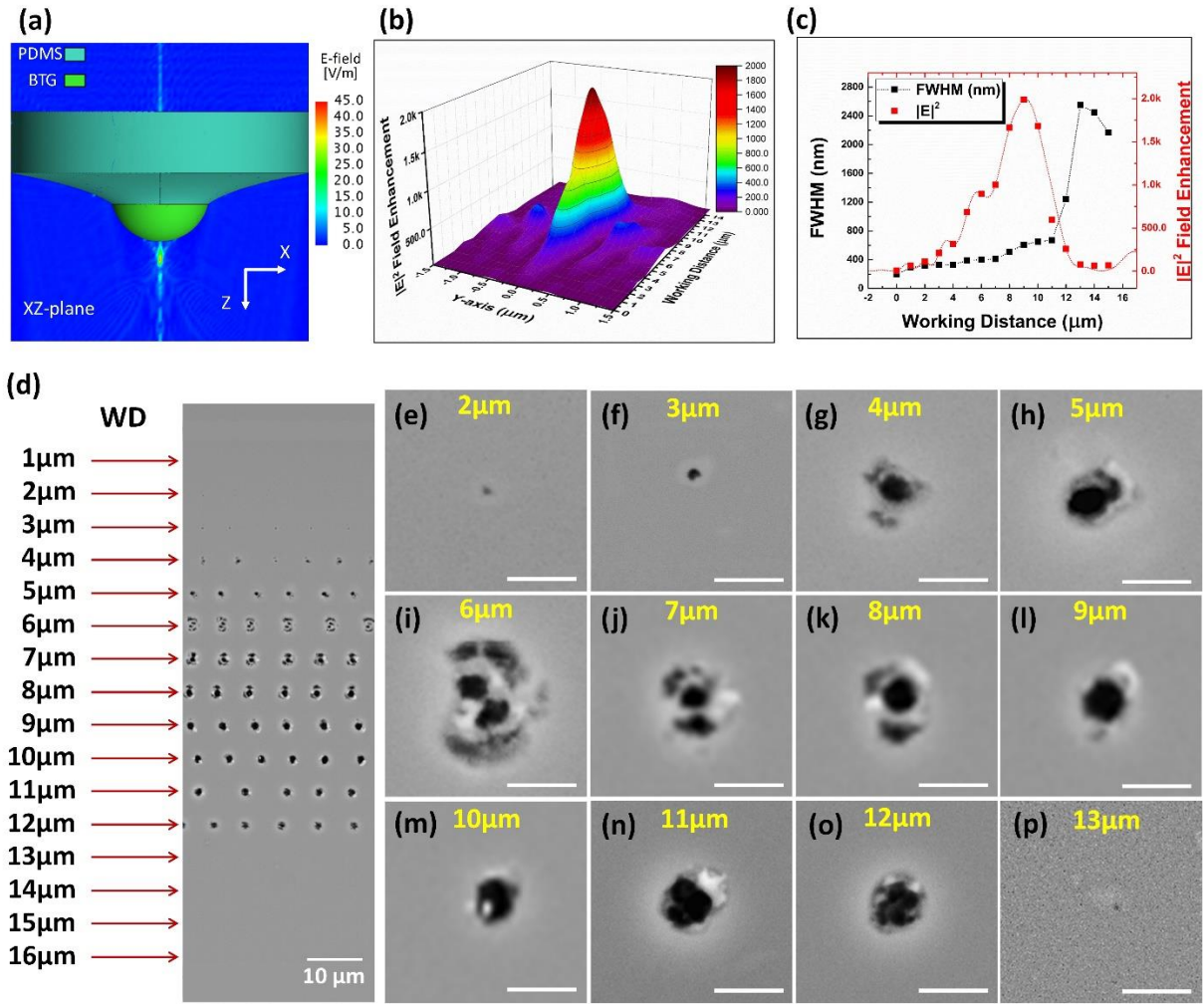

**Figure 2.** Modelling of PCM lens focusing properties and experimental evaluation of femtosecond laser machined feature size against working distance. (a) Electric field distribution of PCM lens at XZ-plane. (b) Field distribution along y-axis and (c) FWHM and $|E|^2$ enhancement at different working distance. (d) Femtosecond laser patterned feature at working distance from 1 to 16 μm, and enlarged SEM images (e)~(p) at 2~13 μm distances, scale bar in (e)~(p) are 2 μm.

To verify the theoretical analysis, experiments were carried out by a femtosecond 800 nm wavelength laser with fluence of 9.5 mJ/(cm)$^2$. A blank silicon wafer was initially elevated to moderately contact with the PCM lens by nano-stage, and then lowered down with 1 μm step until 16 μm apart. Laser was triggered at each working distances and patterned features are shown in Figure 2d. Figure 2e-p are the enlarged SEM images at working distance from 2 μm to 13 μm. There was no feature observed at distances of 1 μm and 14~16 μm, it is because of the low field enhancement or large focal spot in these region resulting in the power density less than the silicon damage threshold, as simulation result showing in Figure 2c. Sub-wavelength features of 230 nm ($\lambda$/3.48) were produced by 2 μm working distance, and it increased to 490 nm ($\lambda$/1.63) at 3 μm apart. Then, the spot size significantly increased to around 1 micron. It is noted that distances at 6, 7 and 8 μm distance, features exhibited not only single hole, but also outer ring patterns simultaneously. This may be determined by the multiple peaks effect, which is consistent with previously discussed theoretical simulation results, shown in Figure 2b. Therefore, high quality laser marking should avoid such circumstances. Meanwhile, the sharpest and roundest hole pattern was found at the distance of 9 μm Figure 2*l*, which can be seen as processing focus point. Furthermore, Figure 2p shows a very shallow and unclear dot generated at 13 μm away from the PCM lens. In general, the experimental results agreed well with the simulation model, and smallest feature size is mainly determined by the PCM lens working distance and laser power density.

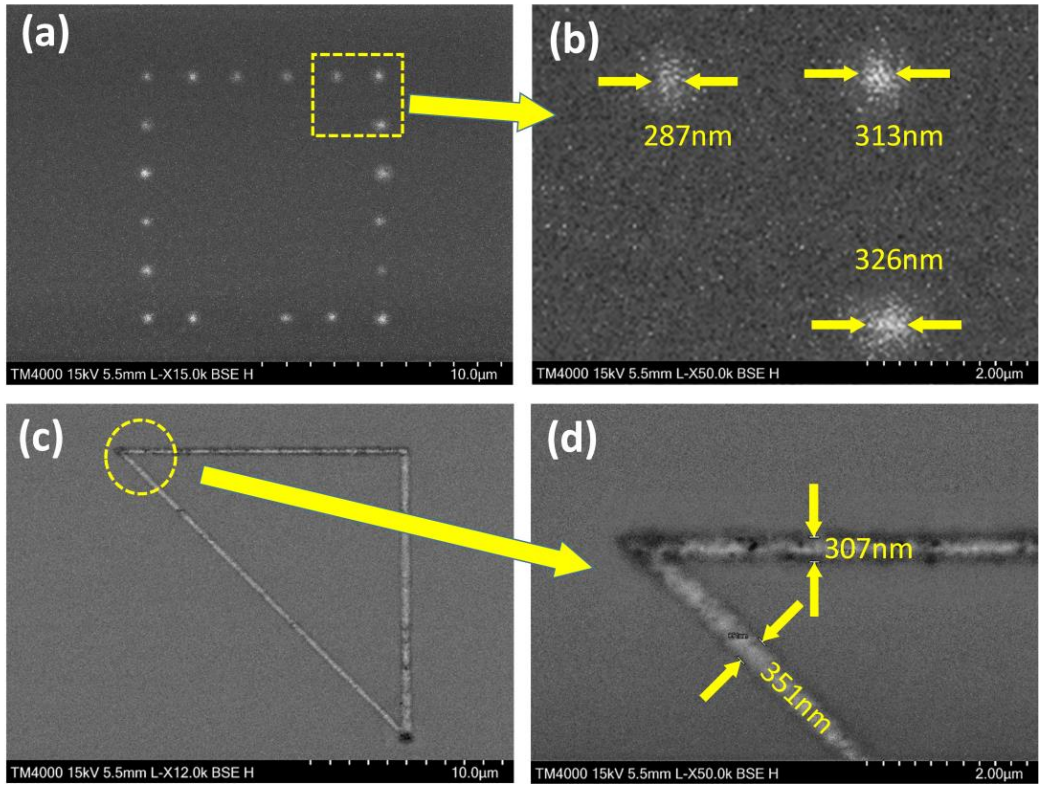

**Figure 3.** Scanning patterning by point-by-point and continuous mode.

In scanning patterning mode, laser was synchronized with the maneuver of piezo nano-stage by in-house developed controlling code, which controlled laser beam on and off while sample moving to realize accurate positioning machining. A blank silicon wafer was scanned over an area in non-contact mode with 2-μm distance under PCM lens, thus avoiding the undesired scratches and distorted pattern from fiction between lens and samples. Here, we presented two scanning modes: point-by-point and continuous mode. In the point-by-point mode, the laser emits a pulse when nano-stage completes a step. Therefore, the pattern outline is formed by dots with certain interval, as shown in Figure 3a and b. On the other hand, Figure 3c and d show the continuously drawing line can be generated by keeping the laser constantly on. The average feature sizes produced by two scanning modes were 309 nm and 329 nm, respectively.

Furthermore, it is important to know that the proposed system is not limited to fabricate simple line and dot patterns, but also flexible to write user-defined complex patterns. Arbitrary pictures can be directly converted into nano-stage recognizable GCS code by self-developed Java program. Therefore, the sample's moving path followed the exact shape from original pictures. The capability of drawing complex arbitrary patterns was evaluated, as shown in Figure 4. Compared to CPLA technique for simple structure in limited area, our system provides a more flexible way in large-area patterning. As a final note, this technique can be further developed by using shorter wavelength laser such as UV laser (355 nm) to improve the patterning resolution down to 100 nm. The PCM lens can naturally being extended to other super-resolution applications including nano-imaging, sensing, trapping and manipulation of nano-objects and samples.

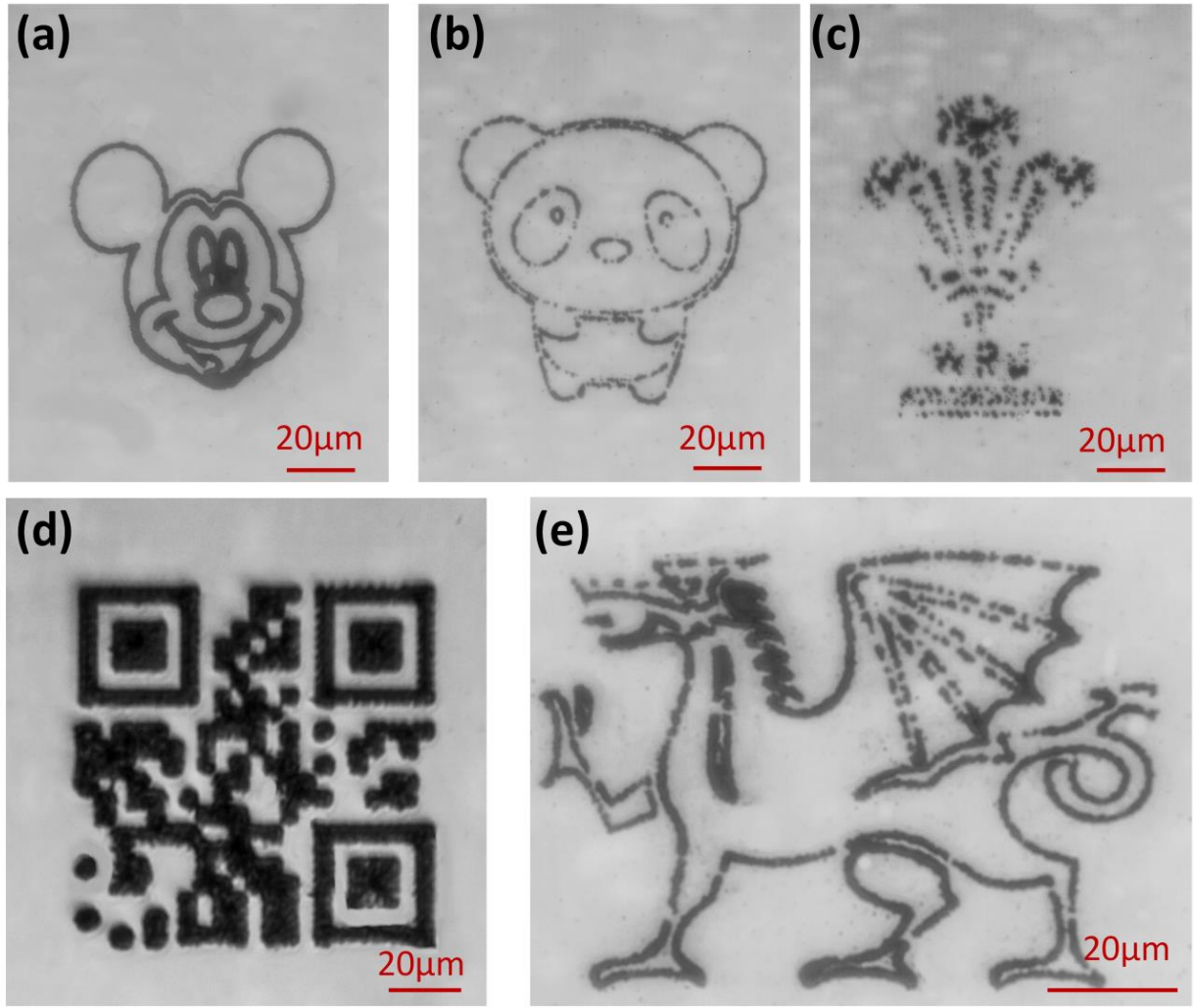

**Figure 4.** Arbitrary patterning on silicon wafer. (a) Mickey mouse, (b) panda, (c) Welsh Rugby Union logo, (d) Bangor University website QR code, (e) Welsh Dragon pattern, scale bars are 20 µm.

**Conclusions**

In summary, we have proposed and demonstrated a new PCM superlens that can be operated in farfield scanning manner to effectively process complex arbitrary patterns in sub-wavelength scale. The effect of non-contact patterning was evaluated theoretically and experimentally at different working distances. The individual features with size of 230 nm ($\lambda/3.48$) was fabricated. The large-area complex structures with arbitrary shapes can easily generated. Meanwhile, the reliability and repeatability were significantly improved compared to other microsphere-based laser fabrication techniques. The development is low-cost and compatible with any existing laser marking system to enhance their pattering resolution.

**ACKNOWLEDGMENT**

Funding support: Center for Photonics Expertise (CPE), European Regional Development Fund (ERDF), Grant No. 81400; Knowledge Economy Skills Scholarship (KESS2, Grant BUK289).